\documentclass[%
 reprint,
 amsmath,amssymb,w
prb,
]{revtex4-1}

\usepackage{graphicx,subfigure}
\usepackage{dcolumn}
\usepackage{bm}
\usepackage{lipsum}

\usepackage{color}
\usepackage{stmaryrd}

\usepackage{amsthm}

\usepackage{nicefrac}
\usepackage{etex}                  
\usepackage{pgfplots}
\usepackage{tikz}
\usepackage{tikz-3dplot}
\usepgfplotslibrary{fillbetween}      
\pgfplotsset{compat=newest}
\pgfplotsset{plot coordinates/math parser=false}
\usetikzlibrary{calc,fadings,shapes.misc,3d,arrows,
 decorations,decorations.pathmorphing,
 decorations.text,shapes.geometric,intersections,
 decorations.markings,patterns,spy,mindmap}

\newcommand{\figref}[1]{Fig.~\ref{#1}}      
\newcommand{\abs}[1]{{\left| #1 \right|}}
\newcommand{\norm}[1]{{\left|\left| #1 \right|\right|}}
\newcommand{\pot}[1]{{\times 10^{#1}}}

\begin{document}

\preprint{APS/123-QED}

\title{Influence of spatial dispersion on surface plasmons, nanoparticles and grating couplers}

\author{Armel Pitelet}
\author{Emmanuel Centeno}
\author{Antoine Moreau}%
 \email{antoine.moreau@uca.fr}
\affiliation{%
Institut Pascal, Universit\'{e} Blaise Pascal,\\
4 avenue Blaise Pascal, \\
63178 Aubi\`ere, France
}%

\author{Nikolai Schmitt}
\author{Claire Scheid}
\affiliation{
Universit\'{e} C\^{o}te d'Azur, Inria, CNRS, LJAD, \\ 
2004 Route des Lucioles, \\
06902 Sophia Antipolis France
}%

\author{Dimitrios Loukrezis}
\author{Herbert De Gersem}
\affiliation{%
Institute for Accelerator Science and Electromagnetic Fields (TEMF), \\ 
Technische Universit\"at Darmstadt, \\ 
64289 Darmstadt, Germany
}%

\author{Cristian Cirac\`i}
\affiliation{
Center for Biomolecular Nanotechnologies, Istituto Italiano di Tecnologia, \\
73010 Arnesano (LE), Italy 
}

\begin{abstract}
Recent experiments have shown that spatial dispersion may have a conspicuous impact on the response of plasmonic structures. This suggests that in some cases the Drude model should be replaced by more advanced descriptions that take spatial dispersion into account, like the hydrodynamic model. Here we show that nonlocality in the metallic response affects surface plasmons propagating at the interface between a metal and a dielectric with high permittivity. As a direct consequence, any nanoparticle with a radius larger than 20 nm can be expected to be sensitive to spatial dispersion whatever its size. The same behavior is expected for a simple metallic grating allowing the excitation of surface plasmons, just as in Wood’s famous experiment. Finally, we carefully set up a procedure to measure the signature of spatial dispersion precisely, leading the way for future experiments. Importantly, our  work suggests that for any plasmonic structure in a high permittivity dielectric, nonlocality should be taken into account.


\end{abstract}


\maketitle

For more than a century now, Drude's model\cite{drude1900}, coupled to Maxwell's equations\cite{maxwell1881treatise}, has been able to describe very accurately the optical response of metals, even for quite extreme geometries\cite{moreau12b,chen2014nanogap,akselrod2014probing,powell2016plasmonic,ayata2017high}. Many advanced theories describing metal-vacuum interfaces\cite{fuchs,feibelman1975microscopic} have been developed during the second half of the twentieth century, especially with the development of Electron Energy Loss Spectroscopy (EELS) which provided experimental data to better ground the theoretical discussions\cite{rocca1995low,rocca1995surface,park2009plasmon}. These new approaches were able to take into account complex phenomena, like spatial dispersion or electron spill-out, and allowed to better understand the success of Drude's model. It turns out in fact that the Drude's theory is the zero-th order approximation of all more advanced descriptions introduced later on. All these studies however, seemed to conclude that spatial dispersion and spill-out have a limited impact\cite{liebsch1993surface,liebsch1993surface1,PhysRevLett.72.788,PhysRevLett.72.789} on surface plasmons (SPs) such that optical experiments were not likely to show any difference from Drude's predictions\cite{boardman82,forstmann86,CHANG2003353}.
Moreover, because most resonances in metallic structures can actually be explained as cavity resonances of some sort for SPs,  it has become widely accepted that plasmonic resonances could be very accurately described by Drude's model. For decades, then, there has not been any urge to adopt advanced descriptions of the response of metals in plasmonics. Only in the case of metallic clusters, due to the extremely small size, spatial dispersion and spill-out were expected to play a significant role, requiring the most advanced descriptions\cite{liebsch1993surface,scholl2012,raza2015nonlocal,esteban2012bridging}.

However, a recent experiment with film-coupled nanoparticles showed that the frequency of the resonance of modes that are localized in small volumes (of the order of 1 nm$^3$) is simply not correctly predicted by Drude's model, whereas the linearized hydrodynamic model\cite{ciraci13}, in its simplest formulation (the Thomas-Fermi approximation) seemed to be accurate enough. This is the case even with an optical excitation and relatively large metallic particles. 
This can be linked to the fact that the small gaps between the nanoparticle and the metal support a gap-plasmon -- a guided mode which is particularly sensitive to spatial dispersion because it has a very large wavevector\cite{moreau13,raza13,david2013perfect,wiener2013nonlocal,toscano2013nonlocal} whatever the frequency. This allows to better understand why small gaps, which are more and more common in plasmonics\cite{akselrod2014probing,moreau12b,lassiter2014third,haffner2015all,nielsen2017giant}, may require more advanced descriptions of the metallic response. Furthermore, this explains why the hydrodynamic model, despite its well documented deficiencies\cite{haberland2013looking}, is probably a good replacement for Drude's model in plasmonics\cite{toscano2015resonance,ciraci2016quantum}: high wavevector plasmonic modes enhance spatial dispersion effects\cite{wiener2013nonlocal,toscano2013nonlocal,david2013perfect}, but not the impact of the spill-out, especially since noble metals present a high extraction work. The spatial dispersion which arises from the repulsion between free electrons inside the metal is taken into account accurately by a hydrodynamic model, which presents in addition the advantage of being relatively easy to implement numerically\cite{toscano2012modified,benedicto2015numerical,schmitt2016dgtd,Schmitt2018}. 

Here we show that the impact of spatial dispersion on the SP propagating at the interface between a metal and a dielectric is enhanced when the refractive index of the dielectric is large. We show, as a direct consequence, that even large nanoparticles can be expected to be sensitive to nonlocality and that, using a grating coupler, it is theoretically possible to estimate the main parameter of the linearized hydrodynamic model, in a situation where other more complex phenomena can be ruled out. We underline that such a well controlled setup differs strongly from previous experiments which all involved chemically synthesized nanoparticles\cite{ciraci2012,scholl2012,ciraci14,raza2013blueshift} whose geometry cannot always be fully controlled.

\section{Surface Plasmon}

In this first part we study the influence of spatial dispersion on a SP propagating along a metal-dielectric interface (especially for dielectrics with 
high refractive indices). A SP can be seen as current loops propagating beneath the surface of a metal. Such a phenomenon is accompanied 
by an electromagnetic field in the metal (with relative permittivity $\varepsilon_{\mathrm{m}}$) and 
in the dielectric (with relative permittivity $\varepsilon_{\mathrm{d}}$) which is transversely evanescent in both media due 
to the fact that its effective index $n_\mathrm{eff} = \frac{k_\mathrm{SP}}{k_0}$ is always 
larger than the refractive index of the dielectric medium. 

\begin{figure}[h]\centering
\includegraphics{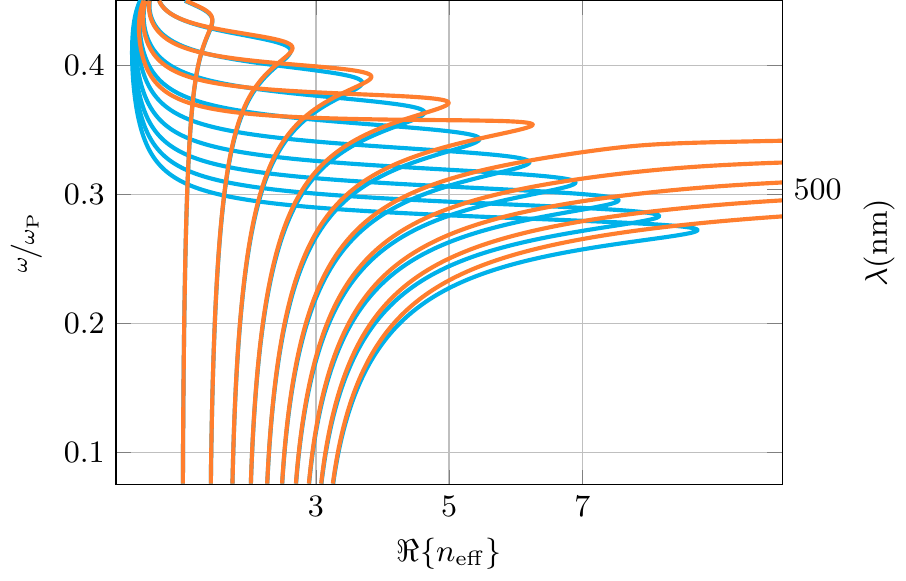}
\caption{Surface plasmon dispersion curves (see relation \eqref{dispersion_nl}) along a single $\mathrm{dielectric-Ag}$ interface, assuming material parameters for silver\cite{rakic98}. The permittivity of the dielectric $\varepsilon_{\mathrm{d}}$ ranges from 1 (most left) to 10 (most right) with a step size 1. 
The blue lines correspond to the local dispersion ($\Omega=0$) and 
the orange ones to nonlocal dispersion ($\Omega \neq 0$).}
\label{multi_diel}
\end{figure}

First, neglecting losses, the dispersion relation can be written as
\begin{equation}
  n_\mathrm{eff} = \sqrt{\frac{\varepsilon_{\mathrm{d}} \varepsilon_{\mathrm{m}}}{\varepsilon_{\mathrm{d}} + \varepsilon_{\mathrm{m}}}},
\end{equation}
where we assumed a simple Drude model $\varepsilon_{\mathrm{m}} = 1 - \frac{\omega_\mathrm{P}^2}{\omega^2}$. 
In this case, the curve of a SP has a horizontal asymptote at
$\omega_\mathrm{SP} = \frac{\omega_\mathrm{P}}{\sqrt{1+\varepsilon_{\mathrm{d}}}}$.
It is however unrealistic to neglect the losses inside the metal because the frequency $\omega_{\mathrm{SP}}$ is usually in a wavelength range, the UV, where the interband transitions make the metal highly lossy. 
As a consequence, a bend-back can be seen on the dispersion curve of the SP, which thus never reaches very high wavevectors (see local dispersion curves in \figref{multi_diel}).

However, when the permittivity of the dielectric increases, the frequency $\omega_{\text{SP}}$ decreases. At the same time, losses due to the interband transitions can be expected to be low enough as to enable a support of high wavevector SPs. Such modes are more likely to be sensitive to spatial dispersion\cite{moreau13}.


In order to take nonlocality into account
we rely on the linearized hydrodynamic model for the free electrons, already introduced in 
previous works \cite{Raza2011,chapuis08,toscano2012modified,ruppin05,ruppin05b,wiener12,ciraci2013effects}. 
The electric current $\mathbf{J}$ inside the metal is linked to the electric field $\mathbf{E}$ by 
\begin{align}
-\beta^2 \nabla(\nabla \cdot \mathbf{J})+\mathbf{\ddot{J}}+ \gamma \mathbf{\dot{J}}=\varepsilon_0 
\omega_\mathrm{P}^2 \mathbf{\dot{E}}, \label{eq_hydro}
\end{align}
where $\omega_\mathrm{P}$ is the plasma frequency, $\varepsilon_0$ the vacuum permittivity, 
$\gamma$ the damping factor and $\beta$ the nonlocal parameter. 
The $\beta$ factor represents the increase of the internal pressure in the electron gas due to exchange interaction and Coulomb repulsion. There are actually several theoretical expressions for this parameter\cite{crouseilles08}. We rely on the experimental data available\cite{ciraci2012,ciraci14}, which consistently  point to a value of $\beta = 1.35\times 10^6 \ \nicefrac{\mathrm{m}}{\mathrm{s}}$. 
Finally, the electric current  inside the metal can always be considered as an effective polarization 
$\mathbf{P}_\mathrm{f}$ due to the free electrons and is then given by $\dot{\mathbf{P}}_\mathrm{f} = \mathbf{J}$. In that framework, the metal can then be described as a nonlocally polarizable medium.

We use accurate material parameters\cite{rakic98} that allow a distinction between the response of the free electrons, that is subject to spatial dispersion, and the response of bound electrons, that can be considered to be purely local\cite{maradudin73}. In the following the metal is always assumed to be silver, which is favorable since silver is less lossy than gold. The total metal polarization reads then $\mathbf{P}=\mathbf{P}_\mathrm{f}+\mathbf{P}_\mathrm{b}$, where $\mathbf{P}_\mathrm{b}=\varepsilon_0\chi_{\mathrm{b}}\mathbf{E}$ with $\chi_{\mathrm{b}}$ being the local susceptibility associated to the bound electrons. In order do obtain $\chi_{\mathrm{b}}$, we fit the experimental data of silver permittivity with a generalized dispersion model based on Pad\'{e} series\cite{lanteri2017} (which is sufficient to perform nonlocal time-domain simulations) and subtract the local Drude contribution $\chi_{\mathrm{f}}$. Throughout this work, we rely on the DIOGENeS \footnote{DIOGENeS, https://diogenes.inria.fr/} Discontinuous Galerkin Time Domain (DGTD) suite\cite{schmitt2016dgtd}. 
Finally we consider additional boundary conditions that are both the most natural (vanishing normal component of the polarization current $\mathbf{J}$ at the metal boundary, thus forbidding free electrons to escape the metal) and the most conservative (they reduce the impact of nonlocal effects)\cite{moreau13}. 

We first study the influence of nonlocality by considering the dispersion 
relation of a SP propagating 
along a perfectly plane dielectric-metal interface. Assuming time dependence of the form $e^{-i\omega t}$ , solving relation \eqref{eq_hydro} for $\mathbf{J}$, injecting the result in Maxwell's equations and considering both classical and additional boundary conditions, the nonlocal dispersion reads
\begin{align}
	\frac{k_\mathrm{m}}{\varepsilon_\mathrm{m}}+\frac{k_\mathrm{d}}{\varepsilon_\mathrm{d}}-i\Omega = 0. \label{dispersion_nl}
\end{align} 
Here, $k_j=\sqrt{\varepsilon_j k_0^2-k_\mathrm{SP}^2}$, $j=\mathrm{m,d}$, are the vertical 
components of the SP wavevector and $\Omega$ is the parameter including $\beta$ and thus taking nonlocality into account. 
Presuming $\Omega=0$ allows to retrieve the usual dispersion relation 
for SPs, the $k_j = k_j'+ik_j''$ being here essentially imaginary ($k_j''>>k_j'$). 
The expression of this parameter is\cite{moreau13}
\begin{equation}
\Omega = \frac{k_\mathrm{SP}^2}{\kappa_{\mathrm{l}}}\left(\frac{1}{\varepsilon_{\mathrm{m}}}-
      \frac{1}{1+\chi_{\mathrm{b}}}
      \right), \label{eq:omega}
\end{equation}
where $\kappa_{\mathrm{l}}^2=k_\mathrm{SP}^{2}+\Big(\frac{\omega_{\mathrm{P}}^{2}}{\beta^{2}}\Big)\Big(\frac{1}{\chi_{\mathrm{f}}}+\frac{1}{1+\chi_{\mathrm{b}}}\Big)$ is the vertical component of the wavevector associated with the longitudinal part of the SP appearing only when considering non-locality\cite{moreau13}. 

As can be seen from \eqref{eq:omega}, the parameter $\Omega$ is roughly proportional 
to the square of $k_{\mathrm{SP}}$, which clearly indicates that the higher the wavevector, the higher the impact of nonlocality. \figref{multi_diel} shows local and nonlocal SP dispersion curves for different values of the dielectric permittivity $\varepsilon_\mathrm{d}$ ranging from 1 to 10. 
We can easily see that, for a fixed frequency, an increasing $\varepsilon_\mathrm{d}$ pushes 
the SP towards higher $n_\mathrm{eff}$, and thus causes the SP to be more  
sensitive to nonlocality. The leftmost curve, obtained using air as dielectric ($\varepsilon_\mathrm{d} = 1$) clearly shows that the effective index $n_\mathrm{eff}$ is too small to show any impact 
of nonlocality. However, after a certain $\varepsilon_\mathrm{d}$ value, the 
characteristic bend-back occurring at the SP frequency $\omega_\mathrm{SP}$ just disappears when spatial dispersion is taken into account.

This corresponds exactly to a recent theoretical study\cite{raza13} which shows 
that an artificial decrease of the metallic losses can induce exactly the same behavior on the dispersion curves of plasmonic guided modes. This occurs when the impact of nonlocality overcomes the influence of the metallic losses. 

The impact of the dielectric's permittivity overcomes the one of losses for two reasons. 
First, as explained above, increasing $\varepsilon_{\mathrm{d}}$ lowers $\omega_\mathrm{SP}$ and thus takes the frequency away from the interband transitions. Additionally, a higher 
dielectric permittivity {\em directly} gives to the SP a higher $n_\mathrm{eff}$ and 
thus enhances the influence of spatial dispersion, leading to a large impact of nonlocality even well below $\omega_\mathrm{SP}$.

We underline that, despite extensive studies on nanoparticles, such a behavior of the SP has seemingly not been reported previously. Straightforwardly, this suggests that nonlocality will have an impact on (i)  the resonances of {\em relatively large} nanoparticles of noble metals (with a diameter well above 20 nm, as they can be considered as resonant cavities for the surface plasmon) and (ii) SP grating couplers very similar to the canonical experiment of Wood\cite{Wood1902} provided the grating is buried in high index dielectrics.

\section{Nanoparticles}

The resonance of large metallic nanoparticles can be interpreted as cavity resonances for the surface mode with a resonance condition\cite{kreibig2013optical} which can be written as
\begin{equation}
    2\pi R = m \frac{\lambda_0}{n_\mathrm{eff}}\label{2pir}
\end{equation}
or simply as $k_\mathrm{SP} = \frac{m}{R}$, where $R$ is the radius of the particle. Such a condition is strictly valid only (i) for a cylinder instead of a sphere and (ii) if the curvature of the particle can be neglected, which is almost never the case.
However, this condition being roughly valid even for spherical nanoparticles instead of cylinders\cite{kreibig2013optical,klar1998surface}, it allows to understand that if the wavevector of the SP is influenced by nonlocality, the resonance frequencies of a nanoparticle should be influenced as well, {\em irrespective of its size}. 

For decades now, the community has actually focused on nanoparticles with a diameter much smaller than 20 nm\cite{scholl2012,raza2013blueshift}, hoping that enhanced nonlocal effects would take place - since spatial dispersion is linked to supplementary pressure terms in the description of the electron gas\cite{crouseilles08,scalora10}. However, with very small nanoparticles (i) other effects like the spill-out\cite{scholl13,esteban2012bridging,teperik2013robust,toscano2015resonance} kick in and (ii) the geometry of the nanoparticles is not well controlled\cite{scholl2012,raza2013blueshift}. Given the relatively poor agreement between experiments and the prediction of the hydrodynamic model\cite{raza2013blueshift,Raza2015}, it seems difficult to consider the hydrodynamic model to be sufficient at such small scales. It may even be a little bit early to introduce further improvements of the hydrodynamic model\cite{mortensen2014generalized} based on these results. The simple analysis above suggests that larger nanoparticles buried in a high permittivity medium could actually be a better setup to test the hydrodynamic model. 

In order to further strengthen this analysis, we used Mie theory\cite{klar1998surface,kreibig2013optical} and adapted the formalism proposed by Ruppin\cite{Ruppin2001} for metallic cylinders, for which formula \ref{2pir} is the most relevant. We use the most realistic material parameters possible\cite{rakic98} and a supplementary boundary condition which can be written\cite{moreau13}

\begin{equation}
    \mathbf{P}_\mathrm{f}.\mathbf{n}=\left(\mathbf{E}-(1+\chi_b)\frac{\beta^2}{\omega_p^2} \nabla (\nabla.\mathbf{E})\right).\mathbf{n}=0
\end{equation}
where $\mathbf{n}$ is the unitary vector normal to the interface. When the field is decomposed on the cylindrical harmonics, the Mie coefficients\cite{Ruppin2001} are
\begin{widetext}
\begin{equation}
        a_n= \frac{\sqrt{\epsilon_d}J_n(k_d R)(J'_n(k_m R)+\alpha_n)-\sqrt{\epsilon_m} J_n(k_m R) J'_n(k_d R)}{\sqrt{\epsilon_m} H'_n(k_d R) J_n(k_m R)-\sqrt{\epsilon_d} H_n(k_d R) (J'_n(k_m R)+\alpha_n)}\label{an}
        \end{equation}
where 
\begin{equation}
    \alpha_n =\frac{\frac{n^2}{k_m R^2} J_n(k_L R) J_n(k_m R)}{
    -k_L J'n(k_L R) + (1+\chi_b)\frac{\beta^2}{\omega_p^2}\left[k_L^2 J_n'''(k_L R) + \frac{k_L^2}{R} J''_n(k_L R) -\frac{k_L}{R^2} J'_n(k_L R)+\frac{2n^2}{R^3} J_n(k_L R)
    \right]
    }
    \end{equation}
    \end{widetext}
with $k_m = \sqrt{\epsilon_m}\,k_0$, $k_d = \sqrt{\epsilon_d}\, k_0$ and 
\begin{equation}
    k_L = \sqrt{-\frac{\omega_p^2}{\beta^2}\left[\frac{1}{1+\chi_b}+\frac{1}{\chi_f}\right]}.
\end{equation}
We then compute the absorption cross-section as 
\begin{equation}C_e = -\frac{2}{k_d R}\sum_{n=-\infty}^\infty \Re (a_n).
\end{equation} 
We underline that when $\alpha_n=0$, the expression \eqref{an} reduces to the regular Mie expression.

\begin{figure}
\includegraphics[width=8.6cm]{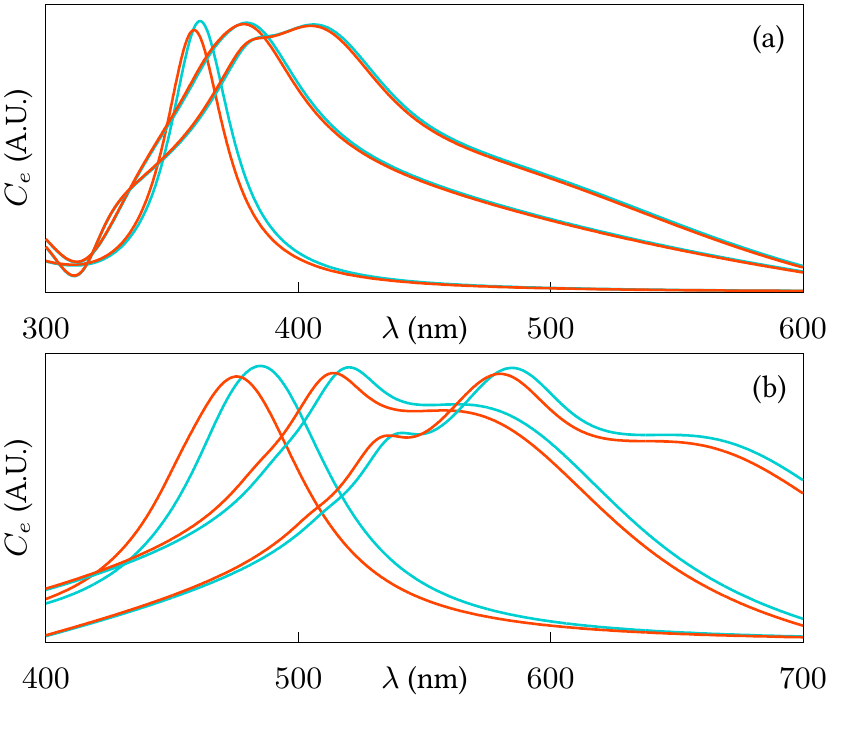}
\caption{Local (blue) and nonlocal (orange) absorption for a cylinder cross-section of {\em radius} (from left to right) $R=10$, $50$, $100$ nm as a function of the impinging wavelength $\lambda$ using \textbf{(a) :} water or \textbf{(b) :} TiO$_2$ as dielectric. The theoretical blueshifts (for the maxima) are from left to right $2.1$, $1.2$, $0.9$ nm on \textbf{(a)} and $9.3$, $6.0$, $4.8$ nm on \textbf{(b)}.}
\label{mie}
\end{figure}

We have computed the local and nonlocal response of nanoparticles with different sizes when they are in water (as is common) or in TiO$_2$ (see Fig. \ref{mie}). The latter is a good choice to enhance the influence of spatial dispersion because of its high refractive index. 
Its permittivity $\varepsilon_{\mathrm{TiO}_2}$ is described by a generalized dispersion model fitted to experimental data corresponding to thin films of $\mathrm{TiO}_2$ grown by atomic layer deposition\cite{siefke2016}. This results in a real part of the refractive index of $\mathrm{TiO}_2$ comprised between 2.25 and 2.5 over a wavelength spectrum ranging from 2000 nm to 400 nm. The extinction coefficient is of the order of $10^{-7}$.

In water, nonlocality has a noticeable impact only for a radius approaching 10 nm, whereas in TiO$_2$ nonlocality tends to blueshift all the resonances of more than 4.8 nm in wavelength even for a radius of 100 nm. We stress here that a high index dielectric is able to sufficiently enhance the magnitude of the nonlocal effects to make it observable on the response of particles/cylinders 5 times larger than the ones usually considered by the community. This should be enough to rule out any other effect like the spill-out and with such a large size, the geometry of the nanoparticles are better controlled - or the nanoparticles could even be probed individually\cite{klar1998surface,Raza2015}.


\section{Grating coupler}

\begin{figure}
\includegraphics[width=8.6cm]{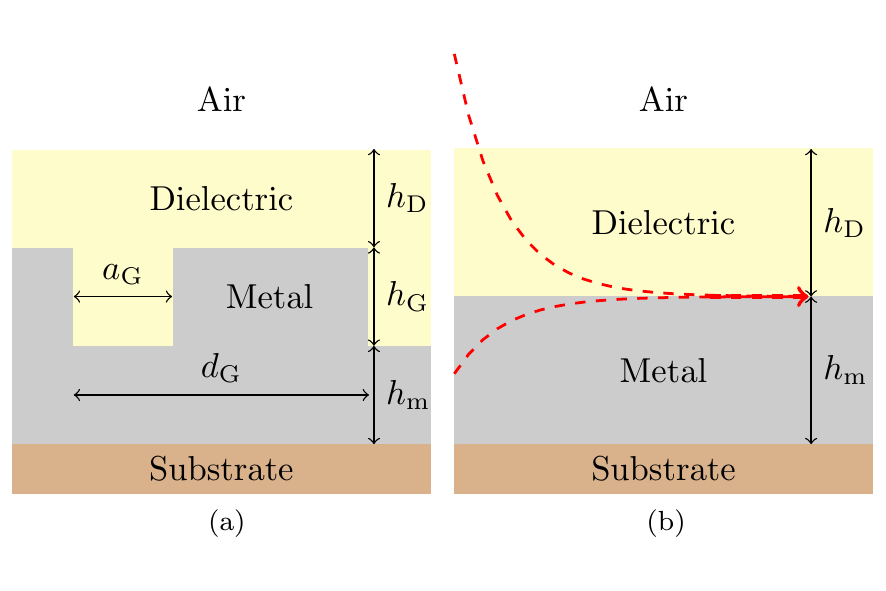}
\caption{\textbf{(a):} Schematic representation of a grating coupler. \textbf{(b):} 
Schematic representation of a SP propagating along a $\mathrm{dielectric-metal}$ interface, taking the finite height of the dielectric into account. The red dashed line illustrates the SP's magnetic 
profile. While the SP propagates along the $\mathrm{dielectric-metal}$ interface, the upper evanescent tail extends into air but the lower one does not extend into the substrate allowing to neglect it in computations.} 
\label{schema}
\end{figure}

We now discuss the structure shown on \figref{schema}(a),  which 
is a simple 1D metallic grating buried in a high index dielectric with infinite extent in 
the lateral directions. Normal incident illumination is assumed from the top  
and we recover the zero-th reflected order for a broadband spectrum of the incident wavelength. 
In such a simple configuration many diffraction orders, including evanescent ones, are excited.

For the $m$-th order, and since we only consider {\em normal incidence} here, the coupling condition to the SP can be simply written as
\begin{equation}
    k_{\mathrm{SP}} = m\,\frac{\lambda}{d_\mathrm{G}},
\end{equation}
where $\lambda$ is the wavelength, $m$ the diffraction order and $d_\mathrm{G}$ the grating pitch. Such a relation is valid only for a very shallow grating when the surface mode can be considered undisturbed. It is equivalent to assuming that the spatial periodicity of the grating is a multiple of the periodicity of the SP. Such a condition is proper to the periodicity of the structure\cite{maystre2012diffraction} and thus is expected to be valid whether spatial dispersion (which will modify $k_{\mathrm{SP}}$) is taken into account or not.  This allows to find the wavelength $\lambda_{\mathrm{c},m}$ for which the grating is able to excite the SP
\begin{align}
 \lambda_{\mathrm{c},m}=\frac{\Re(n_\mathrm{eff})}{m}d_\mathrm{G} \label{couplinglambda},
\end{align}
where $n_\mathrm{eff}=\frac{k_{\mathrm{SP}}}{k_0}$ is the SP effective index.

Each time this condition is satisfied, a dip due to the excitation of a 
SP will appear in the reflectivity. 
Spatial dispersion should cause blueshifts of the resonances with respect to 
a fully local approach because the effective index is always smaller when 
nonlocality is taken into account. We define this blueshift as the positive 
quantity $\Delta\lambda_{\mathrm{c},m}=\lambda_{\mathrm{c},m}^\mathrm{local}-
\lambda_{\mathrm{c},m}^\mathrm{nonlocal} \propto n_\mathrm{eff}^\mathrm{local}-n_\mathrm{eff}^\mathrm{nonlocal}$. 
As the different resonances correspond to different orders being coupled 
to the SP, the corresponding wavevectors will be different and thus 
the impact of spatial dispersion will change from one resonance to another. 
To better understand the impact of nonlocality on the resonances, 
it is crucial to be able to identify them, hence the interest in the relation \eqref{couplinglambda}. We have thus taken the finite thickness 
of the dielectric layer into account by computing the properties of the 
guided mode of the non-corrugated structure (see \figref{schema}(b)) 
using an open-access numerical tool\cite{benedicto2015numerical,defrance16}. 
Aiming for a proper interpretation of the resonances, we have then carefully increased  the depth of the grating, in order to allow a thorough physical discussion (see Fig. \ref{perturbation}). We use Rigorous Coupled Wave Analysis (RCWA)\cite{granet96,lalanne96} for these simulations.


\subsection{Parameters of the grating coupler}

\subsubsection{Materials}

Although we restrict ourselves to silver in the scope of this work, we underline that gold or any other metal 
and even semi-conductors\cite{maack2017size} (claiming an equivalent electronic mean free path\cite{gall2016}) could have
been used and would have led to the same conclusions. Again, in order to observe the largest possible effects, we consider TiO$_2$ as a high permittivity dielectric.

\subsubsection{Thickness of the dielectric and metallic layer} 

We have taken $h_\mathrm{m}=150 \ \mathrm{nm}$ for the thickness of the metallic layer, thus being several times thicker than the skin depth and ensuring that the substrate does not play any significant role.

The thickness of the dielectric $h_\mathrm{D}$ layer is a distinctly more crucial parameter. 
In order to excite a SP with the highest possible wavevector, 
a dielectric of the highest possible thickness would be desirable. However, since the dielectric 
layer is finitely thick, increasing $h_\mathrm{D}$ leads to a higher number of classical 
guided modes. That hinders a clear interpretation of the resonances or renders it even impossible. 
Luckily, since we are considering the excitation of high wavevector guided 
modes that show a fairly low vertical extension, a thickness of $h_\mathrm{D}=85.0$ nm turns 
out to be a good trade-off. No other guided mode than the SP exist for this choice of $h_\mathrm{D}$ in most of the spectrum. 

\begin{figure}[h]
\subfigure[]{
\centering
\includegraphics{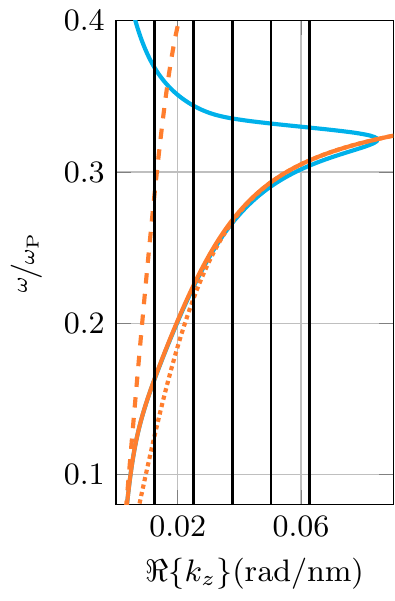}
}
\subfigure[]{
\centering
\includegraphics{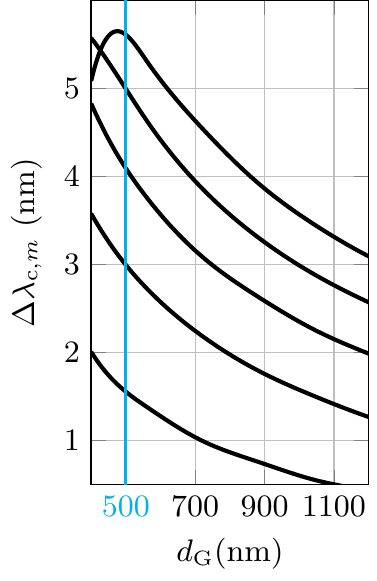}
}
\caption{\textbf{(a):} Dispersion relation of a SP propagating 
at an $\mathrm{air-TiO_2-Ag}$ multilayer. 
Local results are in blue and  nonlocal ones are in orange. 
For comparison, the two dashed lines are the nonlocal dispersion relation respectively 
for an $\mathrm{Air-Ag}$ interface (left) and a $\mathrm{TiO_2-Ag}$ interface (right). 
The black vertical lines indicate the coupling condition given 
by \eqref{couplinglambda} for a grating pitch of $500 \ \mathrm{nm}$  
considering the five first diffraction orders (1 to 5 from left to right). 
\textbf{(b):} Estimation of the blueshift $\Delta\lambda_{c,m}$ 
as a function of the grating pitch $d_\mathrm{G}$. $m$ varies 1 to 5 (bottom to top).
}
\label{mom}
\end{figure}

\figref{mom}(a) shows the dispersion curve for a mode propagating at an air-$\mathrm{TiO}_2$-$\mathrm{Ag}$ interface for $h_\mathrm{D}=85$ nm with 
and without spatial dispersion. Obviously, the difference with the dispersion curve of the SP at a $\mathrm{TiO}_2$-$\mathrm{Ag}$ interface is small. 
The impact of nonlocality is clearly the same and the bend-back disappears.

\subsubsection{Grating pitch}

Using the coupling condition \eqref{couplinglambda} and the local and nonlocal dispersion 
curves as shown in \figref{mom}(a), a raw estimation of the blueshift due to 
spatial dispersion can be made. 
\figref{mom}(b) shows the expected blueshift for different diffraction orders and sweeps over 
the grating periodicity $d_\mathrm{G}$. 
The higher the diffraction order, the higher the wavevector of the excited SP -- which leads to an increased impact of nonlocality. 
We have finally chosen a pitch length $d_\mathrm{G}=500 \ \mathrm{nm}$ according to a maximum of the predicted blueshift of about 5.5 nm for $m=5$. 
Keeping the pitch below the shortest working wavelength guarantees that only the zero-th order propagates, even if the evanescent orders of diffraction are coupled to the SP. In other words, all of the non-reflected light must be absorbed by the guided modes along the structure.

\subsubsection{Groove width and grating thickness}

The determination of the two remaining geometrical parameters, i.e. the groove $a_\mathrm{G}$ and the 
grating thickness $h_\mathrm{G}$, is less straightforward. 
We have to establish a trade-off, such that the excitation of the SP for
different orders of diffraction can be done efficiently without perturbing the guided mode too much. In order to avoid gap plasmons to build up
in the slits\cite{dechaux16}, a large enough $a_\mathrm{G}$ value is needed. We choose $a_\mathrm{G}=\frac{1}{3}d_\mathrm{G}$ here. Starting from $h_\mathrm{G}=2$ nm we have increased the grating depth until an efficient coupling to the SP mode was found - we relied on a pure RCWA method\cite{granet96,lalanne96} to adjust this parameter.
    
\begin{figure}
\includegraphics[width=9cm,height=6cm]{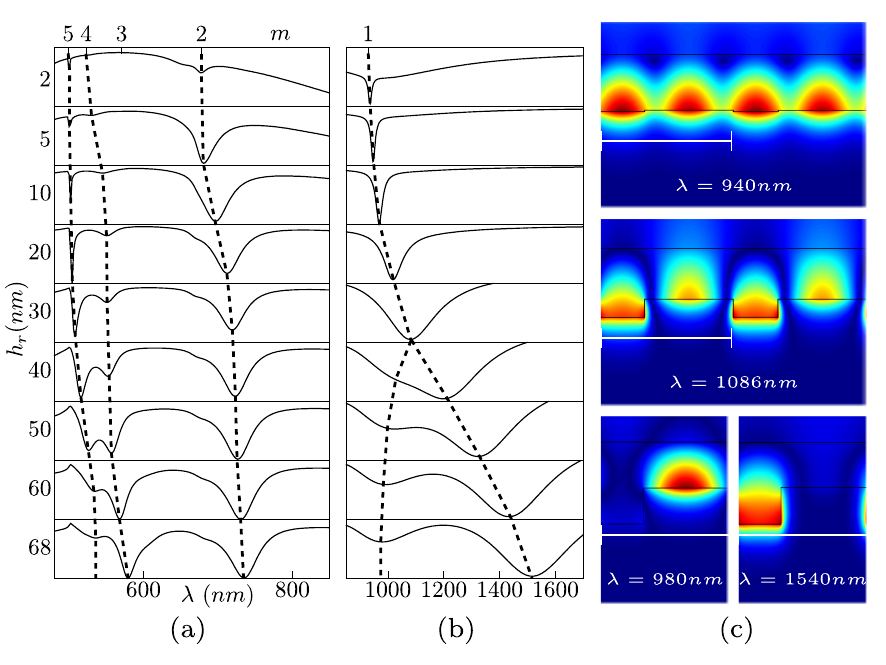}
\caption{\textbf{(a):} Reflectivity of the grating illuminated in normal incidence for different values 
of $h_\mathrm{G}$ computed using a {\em local} RCWA for $\lambda\,\in\,[480,850]$ nm. \textbf{(b):} Reflectivity of the grating for different values 
of $h_\mathrm{G}$ for $\lambda\,\in\,[850,1700]$ nm. Each vertical coordinate system has been chosen to maximize visibility. 
The horizontal one is kept constant from top to bottom. \textbf{(c):} 
Magnetic field amplitudes illustrating the splitting of diffraction order 1 into two sub-orders. The corresponding ($h_\mathrm{G},\lambda$)  couples (in nm) are from top to bottom : (2,940), (30,1086), (68,980) on the left and (68,1540) on the right.}
 \label{perturbation}
\end{figure}

It is easy to associate a diffraction order $m$ for most of the resonances supported by the grating described above (see Fig. \ref{perturbation}). However, starting from almost zero and progressively increasing the height $h_\mathrm{G}$, the resonance that we first attributed to be of order $m=1$ exhibits a splitting (see \figref{perturbation}).

The field maps on \figref{perturbation} clearly show that one of the resonances corresponds to a cavity-like resonance, which is entirely located in the grooves of the grating, i.e it is reflected back and forth horizontally. The other resonances are cavity resonances of SPs that propagate on top of the grating 
and are reflected by the edges of the metal. 
We label those resonances with $1\mathrm{b}$ (where 'b' stands for 'bottom') and $1\mathrm{t}$ (where 't' stands for 'top'), respectively.
Since the thickness of the dielectric is different for both kinds of plasmons, their wavevector cannot be the same, therefore the splitting. We finally chose to take $h_\mathrm{G}=68$ nm.

Were the depth of the grating much larger, the 1t mode would enter a spoof SP regime\cite{pendry2004mimicking,khanikaev2010one,rusina2010theory,kats2011spoof,pors2012localized,ng2013spoof,shen2014ultrathin} where the grating could be considered as an effective medium. In the present case, the conditions to be in this regime\cite{kats2011spoof} are however not fulfilled, as the pitch is roughly one third of the wavelength at most -- for the 1t mode. In addition, the grating is too shallow to allow for spoof SPs to be excited, making 1b and 1t modes more similar to the hybrid SPs evoked in the literature\cite{pendry2004mimicking}. As the following will show, the 1t mode can be considered as insensitive to nonlocality, suggesting that spoof SPs may be equally insensitive.

\subsection{Influence of nonlocality on the grating reflectivity}

\begin{figure*}
\includegraphics{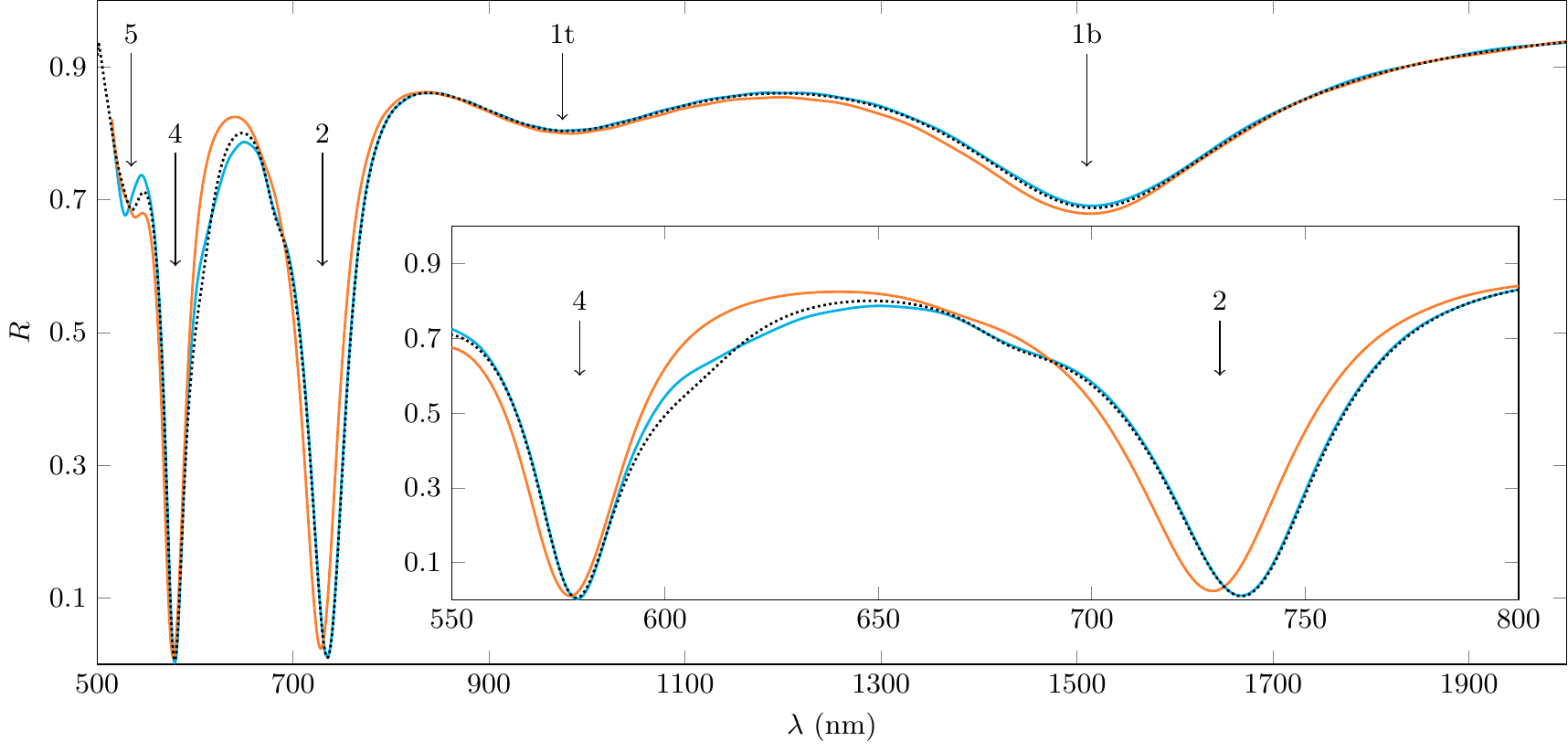}
\caption{Reflectivity $R$ of the whole structure as a function of the wavelength $\lambda$. 
The orange line corresponds to the nonlocal DGTD result, the blue line to the local DGTD result and the black, dotted line to the local RCWA result. 
The annotations $1\mathrm{t}, 1\mathrm{b}, 2, 4$ and $5$ refer to the excitation of the SP with respect to the diffraction order $m$ of \eqref{couplinglambda}. 
The inset plot zooms into the wavelength range $550 - 800$ nm, where the impact of nonlocality is the most prominent.}
\label{fig:mainR}
\end{figure*}

Now that the grating has been designed and its physics is well understood, we use a software taking spatial dispersion into account in order to assess its influence on the reflectivity of the grating. This is necessary, since analytical solutions do not exist anymore and 
we have to rely on a numerical algorithm. Here, we use a DGTD method for the simulations of the grating.
The result is shown in \figref{fig:mainR} for wavelengths ranging from $\lambda=500$ nm to $\lambda=2000$ nm. 
The small discrepancy between local DGTD and RCWA is due to a known problem of the latter in very peculiar conditions\cite{lifeng2011} and occurs only far off the resonances. 
We can clearly identify  5 dips due to SP excitation corresponding to 4 diffraction orders, the first order being split as discussed above. 
The results show a considerable influence of the spatial dispersion, which is significantly stronger than what has been predicted theoretically. 
This can be attributed to the grating itself and to the shift of the resonances towards large wavelengths (and larger wavevectors for the SP). 
The resonances linked to the diffraction orders $m=4$ and $m=2$ (see inset on \figref{fig:mainR}) experience blueshifts of respectively 1.8 nm and 6.5 nm compared to the local prediction. 
Especially for $m=2$ this is clearly higher than the expected shift of 3 nm (see \figref{mom}(b)). 
For $m=4$ the shift is slightly smaller than expected. 
Finally, we can identify two wavelength bands of interest. The first one for $\lambda=[550,800]$ 
nm (see inset) shows the highest difference between local and nonlocal simulations. It 
comprises two blueshifted diffraction orders, and the whole response between the 
two is also clearly affected by nonlocality - making this regime a good choice to estimate the parameter $\beta$. 
The second region of interest is the rightmost part of \figref{fig:mainR} for $\lambda=[800,2000]$ nm. 
It comprises the two suborders $1t$ and $1b$, which are less sensitive to nonlocality. 
For this reason, we believe that this region is not useful to probe nonlocality, but well-suited 
for a geometrical parameter characterization. 
We have to keep in mind that any estimation of $\beta$ relies merely on a comparison between material models. Such comparisons are very sensitive to the geometrical parameters. Since we are trying to measure discrepancies of the order of $1\%$ of the wavelength, we must ensure that nonlocality will not be concealed by uncertainties on local parameters. 

\section{Telemetry and parameter estimation}


In the previous section we have theoretically predicted that the metallic grating that we have designed is sensitive to nonlocality. The goal of the present section is to present the challenges which would have to be faced by experimentalists and to propose a methodology relying on the solution of inverse problems and Uncertainty Quantification (UQ), which could be used to assess realistically the presence of nonlocality and to estimate the crucial $\beta$ parameter.

In order to take into account the fact that, whatever the geometry and the imperfections of the grating, the spectra will be noisy, we have generated an artificial but realistic reflectance spectrum. We use nonlocal simulations and then added a noise whose characteristics are similar to actual experimental spectra provided by experimentalists\cite{jamadi2018edge}. The noise has been chosen with a correlation length of $0.001$ nm and a maximum difference of 0.04 with the original, unperturbed spectrum. We have then tested whether the parameters of the grating and of the model could be retrieved despite this realistic level of noise.

\subsection{Post-fabrication telemetry}

\subsubsection{Grating parameters}

The grating parameters would have to be determined before any dielectrics is deposited. A natural way of determining these geometrical parameters would be to rely on a Scanning Electron Microscope (SEM) or an Atomic Force Microscope (AFM) which would both allow to directly measure the precise parameters for each of the grooves of the grating. However, for a structure without any dielectrics, the impact of nonlocality is negligible - which means that the geometrical parameters can be determined using telemetry, without any assumption on the non local parameter, i.e with a local model. The optical response of a grating actually depends more on {\em average} geometrical parameters\cite{kaliteevski2002disorder}, as the grooves may be all slightly different. Telemetry has the advantage of allowing the determination of these average parameters (especially the period of the grating), which will constitute our geometric model.

We generate an artificial measurement spectrum, that we denote $R_\mathrm{meas}(\lambda)$. Furthermore, it is possible, for given $d_\mathrm{G}$, $a_\mathrm{G}$ and $h_\mathrm{G}$ to compute a theoretical spectrum $R(\lambda)$. We define the distance between the two spectra as
\begin{equation}
    \norm{R_\mathrm{meas}(\lambda) - R(\lambda)}_{L^2(\lambda_1,\lambda_2)}^{2} = \int_{\lambda_1}^{\lambda_2} \left(R_\mathrm{meas}(\lambda) - R(\lambda)\right)^2\mbox{d}\lambda.
\end{equation}
The integration interval $[\lambda_1,\lambda_2]$ will be chosen differently in the following, depending on which parameters have to be retrieved.
We performed multiple optimization runs with different algorithms (particle swarm, pattern search and a derivative-based optimizer combined with a kriging-based meta-model) in order to find the parameters 
($d_{G}$, $a_{G}$ and $h_{G}$) which would produce the minimum distance between the theoretical spectrum and the artificially generated one. We underline that we constrained the optimization to look for geometrical parameters in intervals which would be in accordance with the precision of the etching process. The fabrication tolerances of state-of-the-art nano-processing still lead to a priori tolerances of about $\pm 5 \ \mathrm{nm}$ for the etching process \cite{Langereis2009} and about $\pm 11 \%$ \cite{Chen2015a} for the 
dielectric deposition.

\begin{figure}[h]
\centering
\includegraphics{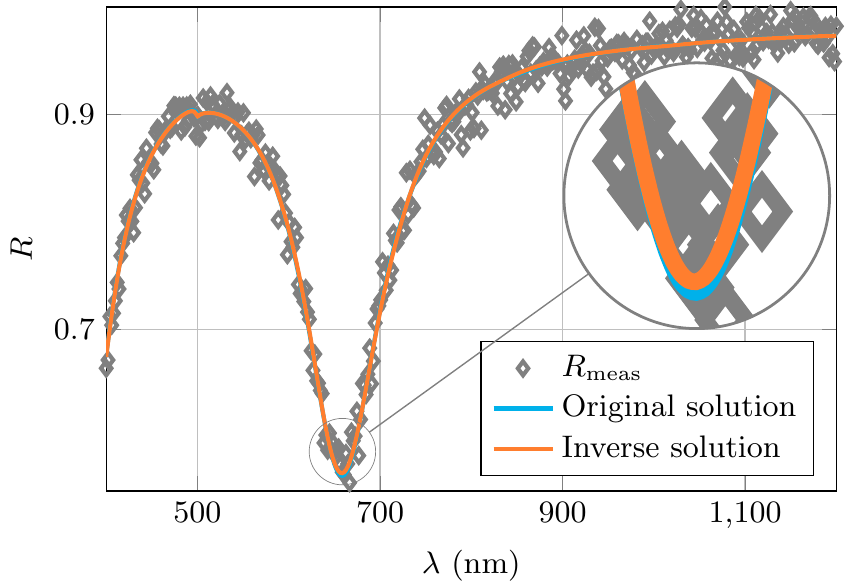}
\caption[Reflectivity spectrum]{Reflectivity spectrum. 
    An artificial white noise has been added to 
    the original spectrum (in blue) obtained for a grating illuminated in normal incidence without any dielectric layer and with $h_\mathrm{G} = 68.0 \ \mathrm{nm}$, 
	$d_\mathrm{G} = 500.00 \ \mathrm{nm}$ and 
    $a_\mathrm{G} = 166.7 \ \mathrm{nm}$. In orange: The result of the optimization.
   This spectrum corresponds to 
    {$h_\mathrm{G} = 68.1 \ \mathrm{nm}$}, 
	{$d_\mathrm{G} = 499.2 \ \mathrm{nm}$} and {$a_\mathrm{G} = 165.4 \ \mathrm{nm}$}. 
	The constraint intervals are chosen to be 
	$h_\mathrm{G} \in [62,73] \ \mathrm{nm}$,  $d_\mathrm{G} \in [495, 505] \ \mathrm{nm}$ and
    $a_\mathrm{G} \in [161, 171] \ \mathrm{nm}$.}
\label{fig:optimizedParameters}
\end{figure}

Considering spectra between 400 nm and 1200 nm to retrieve the parameters, we have found that the derivative-based optimization method performed best, followed by pattern search and particle swarm, the latter seemingly being less appropriate for this type of optimization problem. The geometrical parameters could be retrieved with an excellent accuracy despite the noise (see Fig.\ref{fig:optimizedParameters}). 

\subsubsection{Dielectric thickness}

The dielectric thickness has to be determined by telemetry. While the previous step can be performed with purely local simulations, here nonlocality clearly plays a role. This time, our artificially measured data are generated using the nonlocal spectrum corresponding to the right part of Fig. \ref{fig:mainR} i.e $[\lambda_1,\lambda_2]=[800,2000]$ nm. Since the resonances 1t and 1b are almost insensitive to nonlocality, trying to retrieve $h_\mathrm{D}$ by minimizing the distance between the measured {\em non local} spectrum and a {\em local} in this wavelength range makes sense and actually yields results that are very close to the real value (the retrieved value was $h_\mathrm{D} = 84.8 \ \mathrm{nm}$, with the real value being $h_\mathrm{D, init} = 85 \ \mathrm{nm}$). 


\subsection{Geometric uncertainty vs. nonlocality}

The geometric telemetry has led to the fabricated geometries in Table \ref{tab:parameters}, where, for each parameter $z$ with initial values $z_\mathrm{init}$ used to create $R_\mathrm{meas}$, $\overline z = z_\mathrm{opt}$ is the mean value and $\delta_z = \abs{\overline z - z_\mathrm{init}}$ the maximum deviation.

Given the uncertainties due to the retrieval process (with other means of measuring the geometric parameters these uncertainties would likely be of the same magnitude), we need to be sure that we will be able to distinguish the impact of nonlocality from an unavoidable small error in the determination of the geometric parameters. We use Uncertainty Quantification (UQ) to provide us with answers and, in order to estimate the impact of nonlocality we use, again, a $\beta$ value from the literature\cite{ciraci2012}.

\begin{table}[b]
\caption{Uncertain parameters extracted from telemetry}
\centering\begin{tabular}{llll}
\textbf{Parameter}      &      $\overline z$     &     $\delta_z$     &     \textbf{Units}      \\
\toprule
  $h_{\text{G}}$ & $68.1$ & $0.1$ & (nm) \\
  $d_{\text{G}}$ & $499.2$  & $1.0$ & (nm) \\
  $a_{\text{G}}$ & $165.4$ & $1.5$ & (nm) \\
  $h_{\text{D}}$ & $84.8$  & $0.3$ & (nm) \\ 
\hline
\end{tabular}
\label{tab:parameters}
\end{table}


We recast the geometrical parameters $h_{\text{G}}$, $d_{\text{G}}$, $a_{\text{G}}$ 
and $h_{\text{D}}$ as random variables (RVs) following uniform distributions 
$\mathcal{U}\left[\overline z - \delta_z, \overline z + \delta_z\right]$ (see \ref{tab:parameters}). We then perform an UQ \cite{rcsmith_UQ} study, in order to estimate the expectation value and variance of the reflectivity in two resonance regions corresponding to the diffraction orders 2 and 4. In particular, we consider $25$ wavelengths in the range $[550,600] \ \mathrm{nm}$ ($m=4$) and $50$ wavelengths in the range $[700,800] \ \mathrm{nm}$ ($m=2$).

Since the underlying computational model is a complex one, we rely on black-box UQ methods, i.e. the model and its numerical solvers are used without any modifications.  
In the context of the present work, we employ a spectral method \cite{maitre2010spectral, XiuBook}, in particular the stochastic collocation method \cite{journals/siamrev/BabuskaNT10, journals/siamnum/NobileTW08, journals/siamsc/XiuH05}, taking advantage of a number of factors.
First of all, we deal with a small number of RVs, therefore the costs of the method remain affordable. Moreover, numerical tests indicate that the dependence of the reflectivity upon the RVs is smooth, which is a prerequisite for fast convergence. Finally, we assume that the RVs are mutually independent, which greatly simplifies the method's implementation.

The first step is to choose $M$ different set of values for the geometrical parameters $\mathbf{z} = \left( h_{\text{G}}, d_{\text{G}}, a_{\text{G}}, h_{\text{D}} \right)$, called the collocation points. For each wavelength mentioned above, the reflectance $R(\lambda)$ can be considered as a function $f(z)$ which is approximated by  
\begin{equation}
\label{eq:mtermapprox}
 f\left(\mathbf{z}\right) \approx \sum_{m=1}^M f\left(\mathbf{z}^{(m)}\right) \Psi_m\left(\mathbf{z}\right),
\end{equation}
where  $\mathbf{z}^{(m)}$ are realizations of the random vector (the collocation points) and $\Psi_m$ are multivariate Lagrange polynomials. 
The collocation points are based on univariate Clenshaw-Curtis quadrature nodes and are produced by Smolyak sparse grid rules \cite{bungartz_griebel_2004, Smolyak1963}.
The multivariate polynomials are formatted as products of univariate Lagrange 
polynomials, defined by the corresponding univariate Clenshaw-Curtis nodes.

\begin{figure}
\centering
\includegraphics{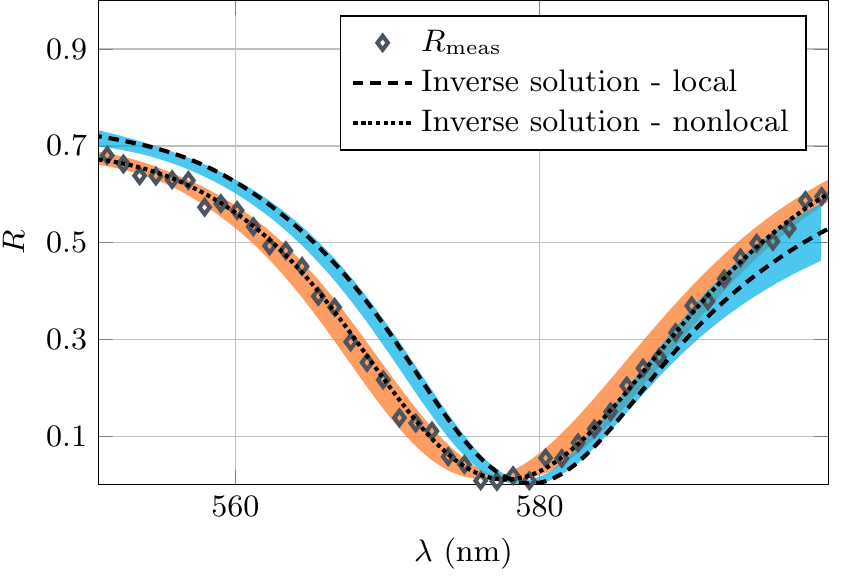}
\caption[Local vs. nonlocal grating]{
    Comparison of the local and nonlocal reflectivity for diffraction order $m=4$. 
    The positions of the local resonances 
    are $579 \ \mathrm{nm}$ and $577 \ \mathrm{nm}$ for the nonlocal one. 
    This leads to a blueshift of almost $2 \ \mathrm{nm}$.
    In blue: the $\mathbb{E}\left[R\right] \pm 2\sigma$ area, being an output of the 
    UQ analysis based on a stochastic collocation method.
    In orange: the min-max intervals of the reflectivity for 
    all interval bound combinations of the geometrical parameters given in Table \ref{tab:parameters}.
}
\label{fig:local_vs_nonlocal_grating_Zoom_res1}
\end{figure}

\begin{figure}
\includegraphics{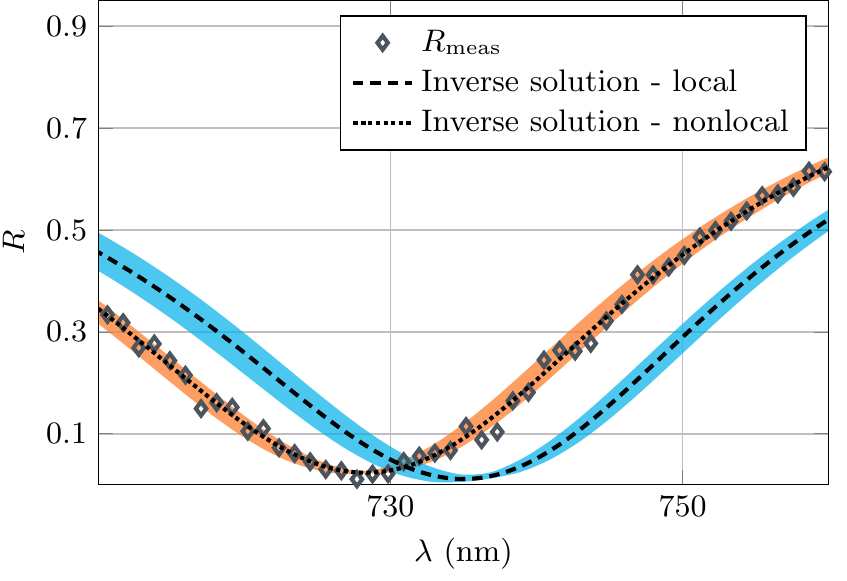}
\caption[Local vs. nonlocal grating]{
    Same as \figref{fig:local_vs_nonlocal_grating_Zoom_res1} for order of diffraction $m=2$. 
    The positions of the local resonances 
    is $735 \ \mathrm{nm}$ and the nonlocal ones at $728 \ \mathrm{nm}$. 
    This leads to a blueshift of almost $7 \ \mathrm{nm}$.
}
\label{fig:local_vs_nonlocal_grating_Zoom_res2}
\end{figure}

The reflectivity's expectation value $\mathbb{E}\left[R\right]$ and variance $\mathbb{V}\left[R\right]$ can now be estimated by post-processing the approximation terms of \eqref{eq:mtermapprox}.
Starting with the definitions
\begin{align}
 \mathbb{E}\left[R\right] &= \int_\Gamma f\left(\mathbf{z}\right) \varrho\left(\mathbf{z}\right) \mathrm{d}\mathbf{z}, \\
 \mathbb{V}\left[R\right] &= \mathbb{E}\left[\left(R - \mathbb{E}\left[R\right]\right)^2\right] = \mathbb{E}\left[R^2\right] - \left(\mathbb{E}\left[R\right]\right)^2,
\end{align} 
where $\varrho(\mathbf{z})$ is the joint probability density function, we approximate the corresponding integrals with the multi-dimensional Gauss quadrature formulas
\begin{align}
\label{eq:E}
 \mathbb{E}\left[R\right] &\approx \sum_{m=1}^M w_m f\left(\mathbf{z}^{(m)}\right), \\
 \mathbb{V}\left[R\right] &\approx \sum_{m=1}^M w_m \left( f\left(\mathbf{z}^{(m)}\right) - \mathbb{E}\left[R\right] \right)^2,
\label{eq:V}
\end{align}
where $w_m$ denote the corresponding quadrature weights.
We use the UQ study results in order to estimate 
$\pm 2\sigma$ intervals around the optimized local 
reflectivity curve, where $\sigma = \sqrt{\mathbb{V}[R]}$ 
refers to the standard deviation. The results corresponding to each 
resonance area are presented in \figref{fig:local_vs_nonlocal_grating_Zoom_res1} and 
\figref{fig:local_vs_nonlocal_grating_Zoom_res2} (in blue), respectively.
Since the nonlocal DGTD simulations are too expensive for an UQ 
of the same kind as we have performed for the local model, i.e. the evaluation of \eqref{eq:E}
 and \eqref{eq:V}, we rely on a min-max study. Here,
min-max represents solver calls for all interval bound combinations (in orange) as depicted 
in the corresponding figures.

  
According to \figref{fig:local_vs_nonlocal_grating_Zoom_res1}, a clear measurement 
of the resonance $m=4$ is almost impossible due to the small difference between the local 
and nonlocal curves. Nevertheless, the second resonance, i.e. $m=2$ (see \figref{fig:local_vs_nonlocal_grating_Zoom_res2})
is significantly more sensitive to nonlocality and stays distinguishable.
For the sake of robustness, we have also performed simulations (data not shown) with the lowest theoretically acceptable value for $\beta=0.85\pot{6}$ m.s$^{-1}$\cite{Hiremath2012}, 
which still guarantees a blueshift of $5 \ \mathrm{nm}$ for the second resonance, i.e. stronger
than the geometric uncertainty.

\subsection{Model calibration}

Knowing the geometry and estimating the impact of nonlocality to be greater than geometric 
uncertainties, we now want to extract the nonlocal material parameter $\beta$. We underline that different theoretical expressions can be found in the literature for this constant\cite{crouseilles08,Hiremath2012,Raza2011}, so that the theoretically acceptable values for $\beta$ may lie between $\beta_\mathrm{min} \approx 0.85\pot{6}$ m.s$^{-1}$ \cite{Hiremath2012} and $\beta_\mathrm{max} \approx 1.4\pot{6}$ m.s$^{-1}$. \cite{Raza2011}. However, the experimental results available so far\cite{ciraci2012,ciraci14} point consistently towards a value close to the upper estimation of $\beta = 1.35\pot{6}$ m.s$^{-1}$.
This underlines how important the determination of $\beta$ can be and explains why we have considered this value so far.

In order to estimate how precise our estimation of $\beta$ could be with the grating setup, we proceed in the same fashion as for the geometric telemetry but we use a wavelength range of $[550,800]$ nm. Using DIOGENeS and DGTD \cite{Schmitt2018}, we find the $\beta$ value which minimizes the distance between $R$ and $R_\mathrm{meas}$.
The geometric size of the structure, in combination with the very small effective wavelengths and the short interaction range of nonlocal effects, which is in the range of several nm at the metallo-dielectric interface, result in a computationally expensive procedure. The solution of the inverse problem can be significantly accelerated by meta-model-based optimization algorithms. We have used the kriging (Gaussian process) meta-model  in combination with a derivative-based optimization implementation of the FAMOSA\cite{Duvigneau} optimization toolbox. We find $\beta_\mathrm{inverse} = 1.385\pot{6}$ m.s$^{-1}$, which is reasonably close to the $\beta_\mathrm{init} = 1.35\pot{6}$ m.s$^{-1}$  (the value used to generate $R_\mathrm{meas}$), indicating that the value of $\beta$ can be retrieved with an error smaller than 10\%.

\section{Conclusion}

We have first shown that, in the framework of the hydrodynamic model, SPs can be sensitive enough to spatial dispersion -- provided that the dielectric considered has a sufficiently high permittivity, like $\mathrm{TiO}_2$. Such a conclusion is in contrast with previous works that suggested that the impact of spatial dispersion could be too difficult to measure optically -- which is only true for an interface between metal and air. 

Since there is a link between such guided modes and the localized resonances of metallic nanoparticles, this lead us to expect an impact of nonlocality on essentially any metallic nanoparticle with a radius much larger than 20 nm, for which the geometry is more likely to be well controlled, buried in a high index medium. Using Mie theory, we estimate the blueshift brought by nonlocality to be at least of 5 nm in wavelength in TiO$_2$, an effect that could potentially be observed experimentally.

Since high wavevector guided modes cannot be excited using prism couplers, we have then studied how the SPs can be excited using a grating coupler. We have shown, using state of the art numerical tools, that such a structure would allow the observation of spatial dispersion by means of blueshifted resonances up to almost 7 nm (around 1\% of the wavelength). Using uncertainty quantification and inverse problem solving, we have identified which resonance precisely could be used to estimate the main parameter of the hydrodynamic model and shown how such an estimation could be made. We underline that such a procedure could well be applied to nanoparticles as well. 

As already evoked in several earlier works\cite{Raza2015,Pitelet2017,pitelet2018} and clearly demonstrated in the present work, any plasmonic structure surrounded by a high refractive index medium like TiO$_2$ will only be accurately described if spatial dispersion is taken into account. 

We believe that, by proposing a structure with realistic parameters and a procedure to carefully estimate the impact of spatial dispersion, this work will pave the way for future experiments that shall give reliable answers to the community on the limits of Drude's model for plasmonics and its potential replacements.


\bibliography{nonlocal}
\end{document}